\documentclass{article}

\usepackage{PRIMEarxiv}

\usepackage[utf8]{inputenc} 
\usepackage[T1]{fontenc}    
\usepackage{hyperref}       
\usepackage{url}            
\usepackage{booktabs}       
\usepackage{amsfonts}       
\usepackage{nicefrac}       
\usepackage{microtype}      
\usepackage{lipsum}
\usepackage{fancyhdr}       
\usepackage{graphicx}  
\usepackage{pgf}
\usepackage{subfigure}
\usepackage{amsmath}
\usepackage{enumitem}

\graphicspath{{media/}}     
\newcommand\inputpgf[2]{{
\let\pgfimageWithoutPath\pgfimage
\renewcommand{\pgfimage}[2][]{\pgfimageWithoutPath[##1]{#1/##2}}
\input{#1/#2}
}}
\title{Testing the efficacy of epidemic testing}
\author{
  Surya Dheeshjith*  \\
  HealthBadge, Inc. \\
  \texttt{surya@healthbadge.org} \\
  \\
  *{Both authors have contributed equally.}
  \And
  Inavamsi Enaganti* \\
  HealthBadge, Inc. \\
  \texttt{inav@healthbadge.org} \\
  \And
  Bud Mishra \\
  New York University \\
  \texttt{mishra@nyu.edu} \\

}

\begin{document}
\maketitle
\begin{abstract}
The cataclysmic contagion based calamity - Covid-19 has shown us a clear need for a comprehensive community based strategy that overcomes the sheer complexity of controlling it and the caveats of current methods. In this regard, as seen in earlier epidemics, testing has always been an integral part of containment policy. But one has to consider the optimality of a testing scheme based on the resultant disease spread in the community and not based on purely increasing testing efficiency.
Therefore, taking a decision is no easy feat and must consider the community utility constrained by its priorities, budget, risks, collateral and abilities which can be encoded into the optimization of the strategy. We thus propose a simple pooling strategy that is easy to customize and practical to implement, unlike other complex and computationally intensive methods.

\end{abstract}


\section{Introduction}

With the advent of Covid-19, complicated by asymptomatic infections, the importance of testing to keep communities safe has become a central topic of investigation (especially when individual members of a community may adopt non compliant behaviour). Many communities faced a lack of testing kits or machines and were not able to conduct a requisite number of tests. Some despite testing were not able to control Covid-19 outbreaks in the community. Some were able to control Covid-19 but with extreme losses in the economic sector which in turn can set off its own disasters. The underlying problem is that traditional systems try to optimize testing as a single standalone entity. Thus the real question is not regarding the required amount of testing to stop the spread but instead how to prioritize one's testing policy to reduce the overall costs while limiting the medical and economic damage to a community. This is no simple problem as every community is vastly different and unique, thus there is a need for a customized strategy.

Let us elaborate with an interesting dilemma. You have to choose between multiple slot machines in a simple gambling game. Table \ref{tab:gambling} shows the cost for playing and distribution of payoff for each slot machine. Assuming you can play only once, what will you choose? 
    
\begin{table}[h]
        \centering
        \caption{Payoff distributions of different slot machines.}
        \begin{tabular}{cccc}
        \toprule
        Cost & Payoff Distribution & Expected Gain\\
        \midrule
             500 & $P(650)=0.5, P(400)=0.5$ & 25\\
             100 & $P(200)=0.5, P(100)=0.5$ & 50\\
             50 & $P(75)=0.5, P(45)=0.5$ & 10 \\
             10 & $P(5)=0.5, P(0)=0.5$ & -5 \\
        \bottomrule
        \end{tabular}
        \label{tab:gambling}
    \end{table}

Observations : 
\begin{enumerate}
        \item Not everyone will choose the same slot machine as it depends on a person's utility, constraints and risk profile.
        \item Observe that paying more does not always translate into a better expected return.
        \item There are natural constraints like a person not having enough money to play some of the slot machines. For example a person with 50 dollars can only play the bottom two games. Alternatively, if constrained to only allow for a loss of up to 5 dollars, only the middle two options are viable.
    \end{enumerate}
   
 These observations draw a compelling parallel towards a community's approach to testing during an epidemic. The priorities and constraints of a community dictate the best course of action for a community. In the case of a very highly infectious epidemic, no amount of testing will suffice while exaggerated testing for simple diseases like the common cold would be a sheer waste of financial resources. Furthermore, traditional systems fail to consider the full picture. They aim for minimum monetary expense for maximum testing efficiency. In that spirit, pool or group testing as originally introduced by Dorfman is gaining prominence [1, 2]. It helps to test more with less and is thus a viable option at a cost of false positives and additional computation or infrastructural setup.\\ 
   
 In this regard, there is a need for a simple system, at the community level that is logistically feasible and fairly inexpensive to implement. We thus propose a community level analysis through a novel but versatile method of a single step random pool testing. Our random pool testing method is easy to implement on existing systems and is also currently being piloted in campuses in India. But to hone a customised yet optimised policy for a community is no easy feat. Thus we additionally introduce Episimmer, an epidemic simulation framework cum decision support system. We built this tool to help community's experiment ideas and test out decisions and simulate their effects on their community. Note that all simulations in this paper have been conducted on Episimmer.

\section{Model Description and Parameters}

\subsection{Environment}

In this article, we consider an agent based analysis to support the effectiveness of our proposed testing system. Despite the heterogeneous nature of real world populations, we consider a homogeneous population of agents that interact along static networks defined by physical proximity. A pair of two agents would be viewed as neighbors in these networks if they spend sufficient time in close physical proximity so that infection can directly transmit from one to the other. Suppose there are $n$ agents which are represented by nodes in the network. We will mainly consider the Erd\H{o}s-R\'{e}nyi (ER) random graph G(n,p) with the connectivity parameter $p$ on these $n$ nodes.

\subsection{Disease Model}

In epidemiological literature, several agent-based stochastic models have been proposed and studied to capture different aspects of the infection spreading mechanism. Examples include (i) SI (susceptible-infected), (ii) SIR (susceptible-infected-recovered), (iii) SIS  (susceptible-infected-susceptible), (iv) SIRS (susceptible-infected-recovered-susceptible), (v) SEYAR (susceptible-exposed-symptomatic-asymptomatic-recovered) epidemic models and their variants [3]. Although this paper considers the simple SIR model, it can be easily extended to any of the more complex epidemiological models. In the SIR model, each person has one of the three possible states -- susceptible ($S$), infected ($I$), and recovered ($R$). An individual moves from S compartment to I compartment at rate $\beta$ times the number of $I$ neighbors. Finally, people move from I compartment to R compartment at rate $\gamma$ through recovery. At a given point in time, the proportion of infected is called the prevalence denoted by $\lambda$ where $0<\lambda<1$.

\subsection{ Testing Policy}

Testing is the process of evaluating the presence of disease in an individual. Testing is done to identify infected individuals and mitigate disease spread to the rest of the susceptible population. We consider the simplest testing process with a binary result which can either be positive (infected with disease) or negative (not infected). Diagnostic tests are not perfect and thus have an associated sensitivity (True positive rate) and specificity (True negative rate) to characterise their veracity. We denote the false positive rate and false negative rate by $\mu_{+}$ and $\mu_{-}$ respectively. Test results may not be immediate and have a latency. The period of time taken to process the test sample and return a result is known as the turnaround time $T_{T}$. Practically, in a community, testing may not occur constantly. Usually a specific day or feasible duration is allocated for testing. These periods of testing can be further categorised by the number of tests conducted in the period ($F$) and the gap between consecutive periods ($T_{G}$). Testing needs to be followed up by an intervention policy to stop spread. These typically include contact-tracing, isolation, treatment or restriction. We consider a period of restriction ($T_{R}$) post testing positive.\\

As seen in the early days of Covid-19, there is a huge lack of tests/testing machines. This limitation drives forth the need to increase testing efficiency while constrained by the number of tests one can conduct. It motivated us to pool or group testing as originally introduced by Robert Dorfman. He aimed to reduce the expected number of tests needed to identify all infected by grouping agents or individuals together while testing. Pooling results in a reduced latency or turnaround time while also reducing the cost of testing an individual. There are also disadvantages to general pool testing but shall be addressed by our novel testing method.

\subsection{Simulation}
We consider a finite horizon with a fixed time frame. Throughout this paper, in order to reduce variance in this complex non-deterministic system with largely inherent stochasticity or randomness, all shown results are averaged over multiple instances of the simulation. The number of instances were determined by the convergence of the final result. Although variance can be introduced as a risk metric, this is not in the scope of this paper.

\section{Our Pool testing method}
Pool testing is a relatively old idea [1] and there exists a large number of pooling methods today [4, 5, 6]. Many of these ideas have their shortcomings and thus have been subjected to criticism. Multistep methods like Dorfman pooling and its variants tend to increase the logistic complexity and require multiple rounds of testing. Combinatorial ideas which use compressed sensing [7, 8] and encoding methods [9], increase computation along with other difficulties in their practical realization (Additional logistic or technical training). Furthermore, these methods are constrained by bounds on prevalence, inherent false positive and false negative rate of the testing process. Also these methods look towards improving efficiency and do not factor in the disease dynamics and the environment as a whole. Thus there is a calling for a new and improved method.

\begin{figure}[ht]
  \begin{center}
    \subfigure[]{\includegraphics[scale=0.49]{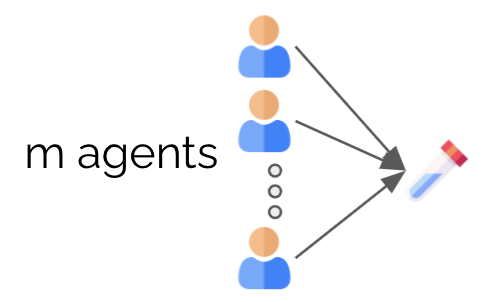}}
    \hspace{1in}
    \subfigure[]{\includegraphics[scale=0.49]{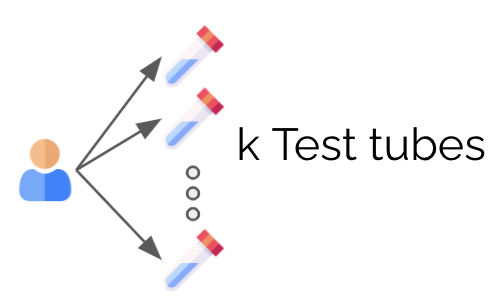}}\\
  \end{center}
  \caption{\textbf{}}
  \label{fig:naptntpa}
\end{figure}

In this section, we describe our simple single step pooling method. We define our pooling strategy with two parameters - Number of Agents per test tube ($m$) and Number of Test tubes per Agent ($k$). As shown in Figure \ref{fig:naptntpa}, $m$ refers to the maximum pool size or the number of agents present in a pool (test tube / sample collection container), and $k$ refers to the number of distinct pools the agent can put their sample into. We can thus represent a specific strategy by the tuple ($m, k$). (1,1) represents traditional testing where a each test tube contains one and only one agent's sample. 

We consider $r$ agents testing in $F$ pools forming a regular bipartite graph with the agent and test tube components having degree $m$ and $k$ respectively. We assume that $0\sim\mu_{-}<<1$ and thus negative on any pool is considered a negative certificate for the agent. If all pools are positive then the agent is given a positive certificate. Agents with positive certificates are restricted accordingly. Table \ref{tab:math} shows the probability of getting a correct or wring certificate. This can be adjusted with an appropriate pooling strategy.

Given an infected agent, 

\begin{align*}
    Probability\ that\ an\ infected\ agent\ \\ gets\ a\ positive\ certificate &= Probability\ of\ all\ pools\ agent\ tests\ in\ is\ positive\ \\
    &= P(Infected\ Pool\ testing\ positive)^k\\
    &= (1-\mu_{-})^k \\
    Probability\ that\ an\ infected\ agent\ \\ gets\ a\ negative\ certificate &= 1-(1-\mu_{-})^k
\end{align*}
Given a non-infected agent. If we assume a large number of people and number of pools,

\begin{align*}
    Probability\ that\ a\ healthy\ agent\ \\gets\ a\ positive\ certificate &= Probability\ of\ all\ pools\ agent\ tests\ in\ is\ positive\\
    &\sim P(Pool\ testing\ positive)^k\\
    &\sim \Big ( P(remaining\ pool\ has\ infected) \cdot (1-\mu_{-})\\ &\ \ \ \ + P(remaining\ pool\ has\ no\ infected) \cdot \mu_{+} \Big )^k\\
    &\sim \Big ( \big (1-(1-\frac{n \lambda}{n-1})^{m-1} \big ) \cdot (1-\mu_{-}) +  (1-\frac{n \lambda}{n-1})^{m-1} \cdot \mu_{+} \Big )^k  \\
    Probability\ that\ a\ healthy\ agent
    \\gets\ a\ negative\ certificate &= 1 - \Big ( \big (1-(1-\frac{n \lambda}{n-1})^{m-1} \big ) \cdot (1-\mu_{-}) +  (1-\frac{n \lambda}{n-1})^{m-1} \cdot \mu_{+} \Big )^k
\end{align*}
If we assume a large enough $n$, then $n/(n-1) \sim 1$ resulting in the values as shown in Table \ref{tab:math}.


\begin{table}[h]
        \centering
        \caption{\textbf{Probability of getting a positive or negative certificate with a (k,m) pooling strategy.} Given the state of an individual, $\mu^+, \mu^-$ and external prevalence $\lambda$.}
        \begin{tabular}{|c|c|c|}
        \toprule
         & Non-infected & Infected\\
        \midrule
             P(Positive certificate) & $\Big ((1-\mu_{-}) - (1-\lambda)^{m-1} \cdot (1 - \mu_{+} - \mu_{-})\Big )^k$ & $(1-\mu_{-})^k$\\
             \midrule
             P(Negative certificate) & $1-\Big ( (1-\mu_{-}) - (1-\lambda)^{m-1} \cdot (1 - \mu_{+} - \mu_{-})\Big )^k$ &$1-(1-\mu_{-})^k$\\
            
        \bottomrule
        \end{tabular}
        \label{tab:math}
    \end{table}

\begin{figure}[ht]
  \centering
    \resizebox{1.00\textwidth}{!}{\input{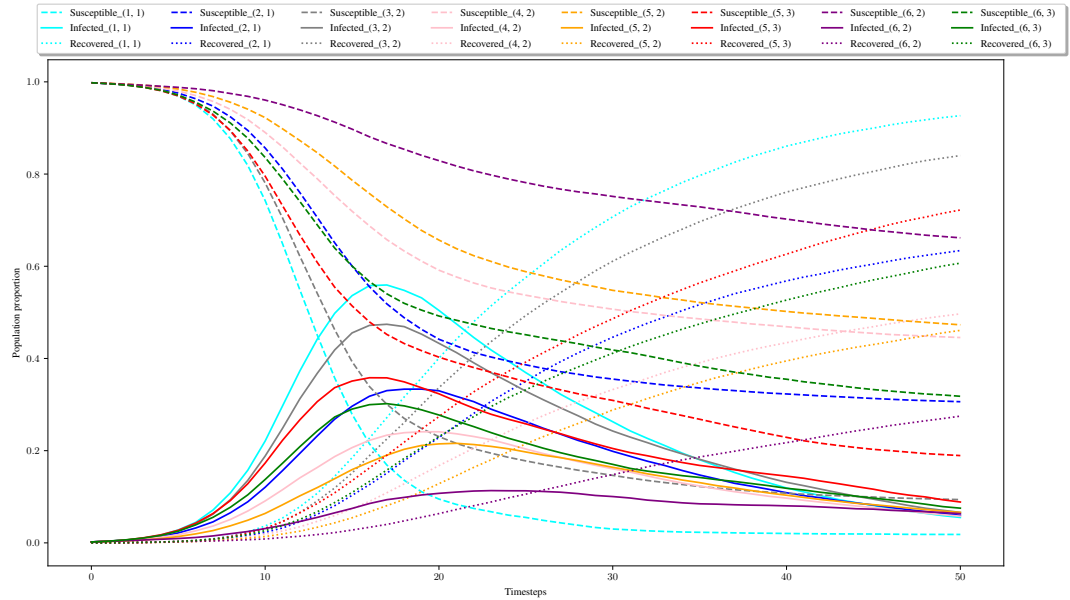}}\\
  \caption{\textbf{Comparison of disease spread with different pool testing strategies followed by restriction.} This time series showcases the proportions of states (S,I,R) in the population at a given time. Each colour represents a given pooling strategy while the connector symbols represent a state each. Observe that pooling strategies control the disease in the decreasing ratio of $m/k$(number of agents testing / total number of pools). Thus (6,2) controls the disease best while the pooling strategy (1,1) results in the whole population getting infected quickly.}
  \label{fig:multiSIR}
\end{figure}

This pooling method was conceived to be as versatile and adaptive as possible keeping in mind the practical ease with which it can be implemented in the real-world. Figure \ref{fig:multiSIR} shows various strategies and their effects on a simulated community. 

\begin{figure}[ht]
  \begin{center} 
    \subfigure[]{\includegraphics[scale=0.33]{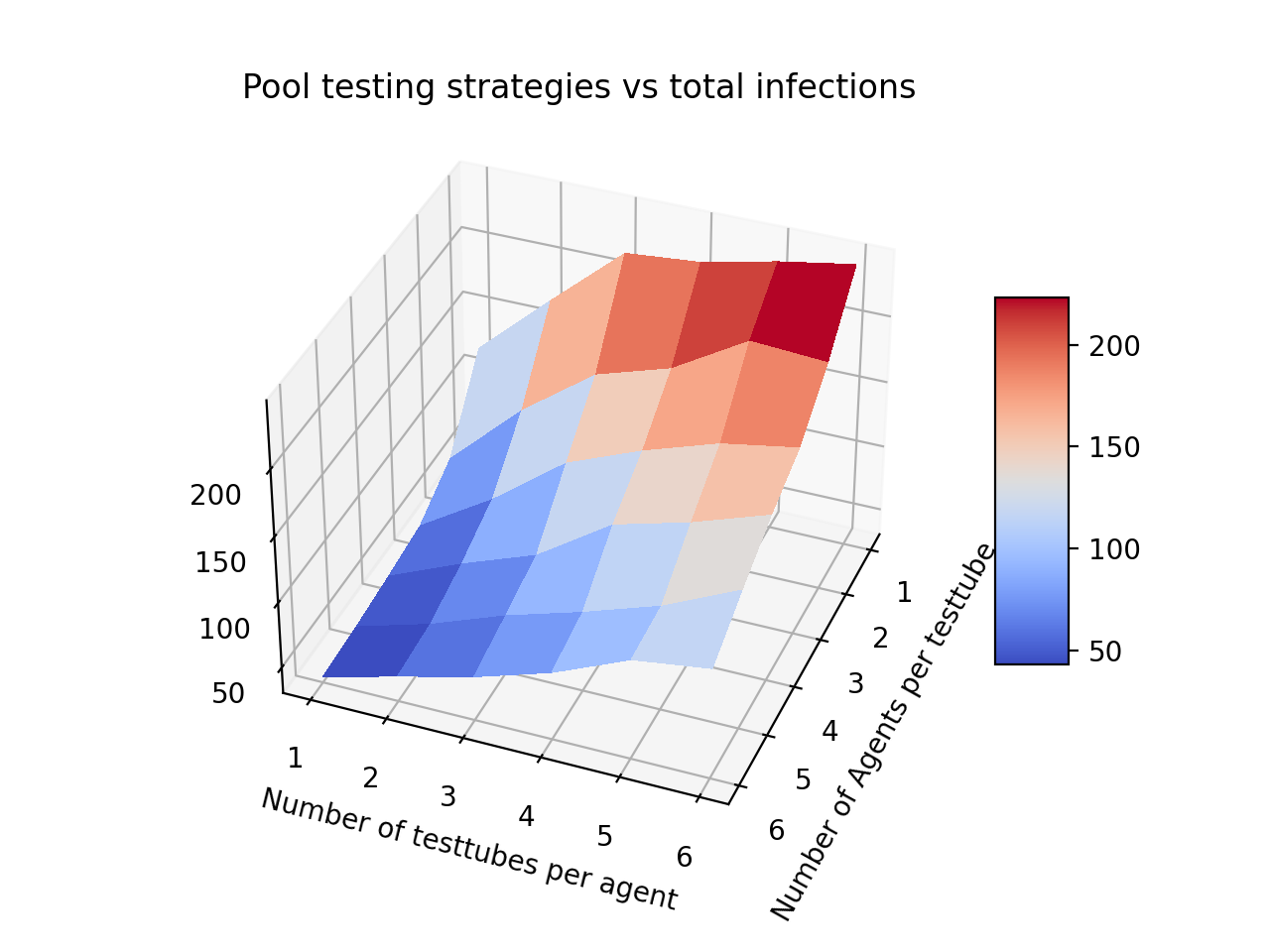}}
    \subfigure[]{\includegraphics[scale=0.33]{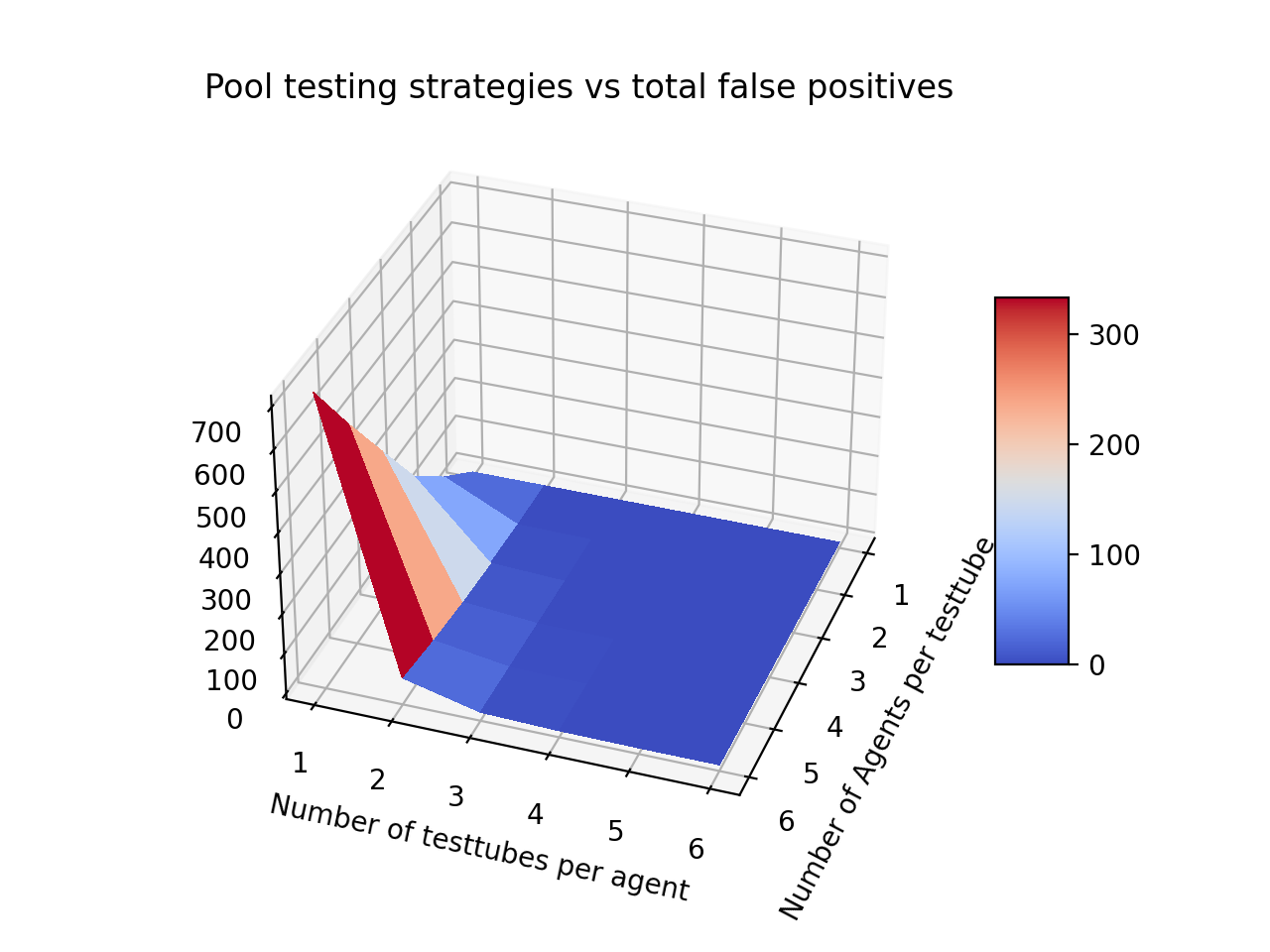}}
    \subfigure[]{\includegraphics[scale=0.33]{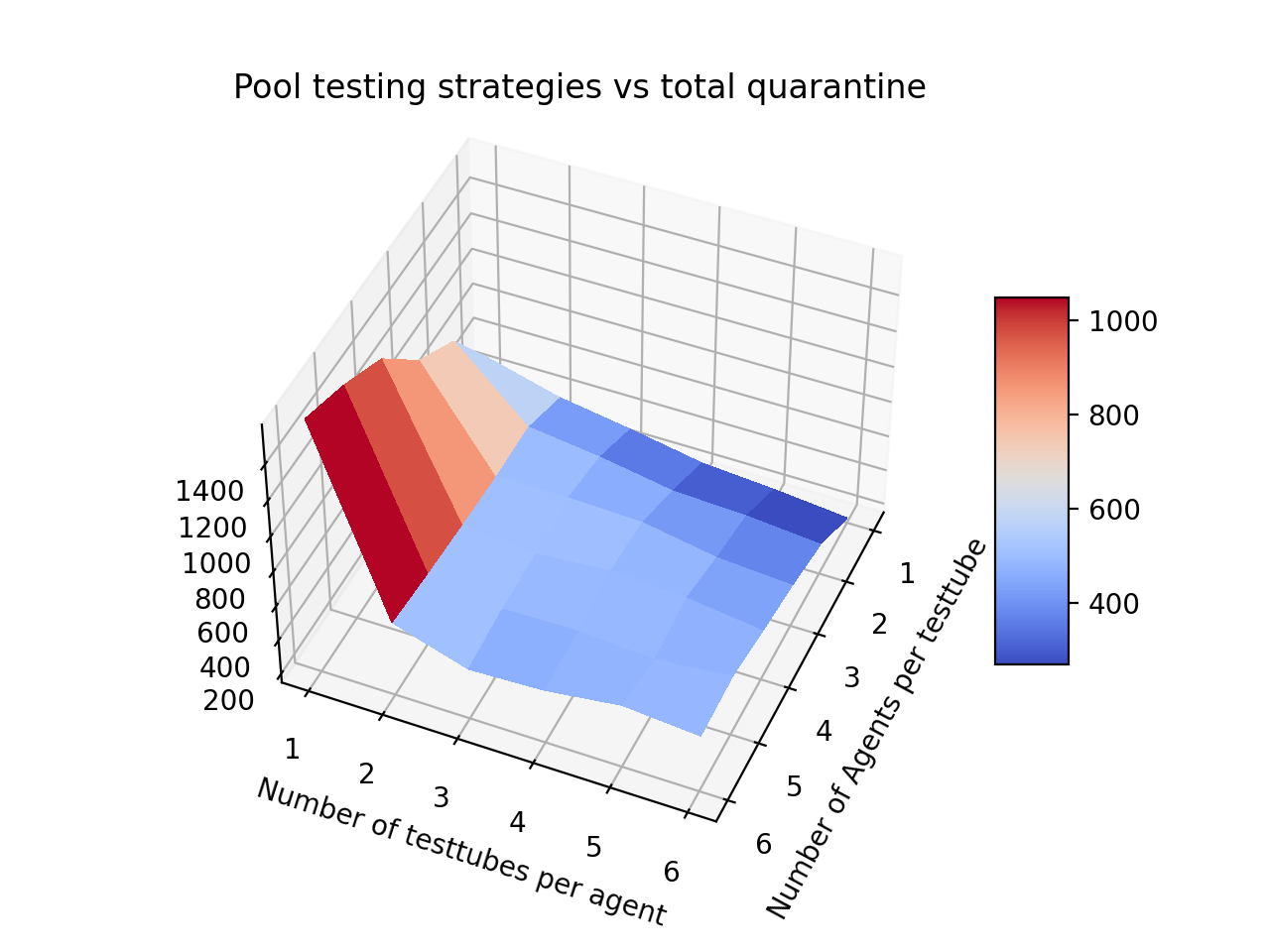}}
  \end{center}
  \caption{\textbf{Visualising tradeoff in effects of various pool testing strategies.} The simulations have been run on a G(1000,0.01) ER graph with $\beta=0.02$, $\gamma=0.12$. Subfigures (a),(b) and (c) represent surface plots of total infections, total false positives and cumulative quarantined days across all agents given a 2-dimensional pool testing strategy respectively. The sub figures depict the tradeoff of different quantities for a given strategy.}
  \label{fig:3D_plot1}
\end{figure}

Observe how the effects of different strategies vary drastically because the chosen testing strategy affects the disease spread by restricting those identified as positive. This change in prevalence affects the results of the next testing period as shown in Table \ref{tab:math}. To better understand this we do a comparative simulation study. Table \ref{tab:tradeoff} quantifies these effects for an environment where we conduct 90 tests on alternate days. The table portrays the tradeoff of total infections, false positives and quarantined days for given strategies. Figure \ref{fig:3D_plot1} visualises the tradeoffs represented by the table.

\begin{table}[ht]
    \caption{\textbf{Results of various pool testing strategies.} The simulations have been run on a G(1000,0.01) ER graph with $\beta=0.02$, $\gamma=0.12$.}
    \centering
    \begin{tabular}{cccccccccc}
        \toprule
        (napt,ntpa) & (1,1) & (2,1) & (3,2) & (4,2) & (4,3) & (5,2) & (5,3) & (6,2) & (6,3)\\
        \midrule
         Agents/day & 90 & 180 & 135 & 180 & 120 & 225 & 150 & 270 & 180 \\
         Total Infection & 108.63 & 64.97 & 89.4 & 68.27 & 93.37 & 55.37 & 75.0 & 52.4 & 64.27 \\
         Total False Positives & 0.0 & 84.37 & 11.0 & 18.43 & 2.47 & 35.4 & 4.77 & 59.93 & 8.1 \\
         Total Quarantined Days & 409.73 & 811.93 & 545.4 & 509.6 & 475.3 & 546.73 & 468.23 & 612.93 & 467.37 \\
         \bottomrule
    \end{tabular}
    \label{tab:tradeoff}
\end{table}

In Figure \ref{fig:3D_plot1}, observe that the infections are directly proportional to $m/k$, which is a result of testing more agents. Observe that most of these strategies maintain a low number of false positives. This analysis shows the practical feasibility of these strategies. Observe the nature of total quarantined days. This interesting shape arises because a high $m/k$ ratio tests more agents early on resulting in a lower epidemic size and correspondingly a lower total quarantine. \\

This interesting tradeoff signals a choice that communities have to make. The choice of an appropriate strategy is determined by the priorities of the community. This result brings forth the question of the best strategy for a community. We address this formally in the following section.

\section{Optimal Testing}

\subsection{Cost Structure}

For every policy the community seeks to implement, there is an associated cost borne by the community. Here, cost does not imply the mere monetary expense but a total encoding of all effects. This cost structure encodes the priorities, expenses and constraints of a community. \\
\begin{itemize}
    \item Priorities :  Every community is unique in their needs, vulnerabilities and abilities which determine their priorities. Quarantine for an affluent community with savings is not as harsh as quarantine of a less fortunate community that lives on a day to day wage. Furthermore communities differ in inherent strengths and weaknesses. For example a nursing home is much more likely to result in deaths as compared to an undergrad college with the same of number of cases.
        
    \item Expenses : The expenses associated with a community depends on many other factors including the demographic composition, socio-economic factors, healthcare infrastructure, the level of awareness, perception, and sensitivity of people. At a basic level the GDP, technological progress, infrastructure and resources of a country determine various expenses. Even at a local level expenses can vary across neighbourhoods due to availability of labour, demand and supply of resources and transport. An example is that of reagent, testing personnel, quarantine and machine expenses. Among the available testing machines, their cost monotonically decreases with false positive rate, false negative rate, and turnaround time. 
    \item Constraints : A variety of factors like Infrastructure, Availability of resources, Demand and Supply, Government Regulations constrain the possible policies a community can implement. Constraints could include a minimum number of infections, deaths, quarantine days and could also include bounds on the total number of false positives and false negatives in a certain period.
\end{itemize}

We consider a cost structure that is composed of the following four components:

\begin{enumerate}
\item $C_1$ : cost for a test -- the cost of conducting a single test and getting results. This parameter includes the reagent, logistic and technological costs. 
\item $C_2$ : cost for a false positive result -- a false positive result stems from two factors. One factor is the inherent accuracy and sensitivity of the testing equipment. This parameter also encodes the chances of sample being contaminated or manual errors during testing. The second factor is the false positives which arise due to the pool testing mechanism. In this paper we consider the cost for one agent getting a false result. Note that the false negative cost is encoded in the total infections, as letting an infected person go unquarantined results in the risk of more infections.
\item $C_3$ : cost for quarantining or restricting a single agent for a day -- agents will be restricted and thus will not be able to take part in economic activities involving physical interactions. This cost encodes the economic loss. 
\item $C_4$ : cost incurred by an infected agent for a day -- this cost stems from the fact that the infected people with health cost due to hospitalizations, health hazards, and fatalities. This cost encodes the medical and logistical costs for an infected person. 
\end{enumerate}

Another main component is the fixed costs that includes setting up the pooling system and other processes like information dissemination. Since it is fixed, we formalise our cost structure with only the four components as a tuple $(C_1,C_2,C_3,C_4)$.

\subsection{Cost Function}

Given a policy, the cost function returns the total cost borne by the community, which is determined by the cost structure. 

We consider the cost function as the linear combination of the cost components. 

\begin{equation}
    \mathcal C = n_1 C_1 + n_2 C_2 + n_3 C_3 + n_4 C_4 
\end{equation}

Where, $n_1$, $n_2$, $n_3$ and $n_4$ are the total number of tests, false positives, quarantined days and infections respectively.\\

We have considered linear cost components but this may be appropriately adjusted to capture the community's utility. For example a community which shuns excessive deaths could penalise the cost function with a quadratic component in the number of deaths.\\

\textbf{Goal} : Find the optimal pooling strategy that minimizes the cost function.

\subsection{Dependent Parameters}

\begin{table}[ht]
    \caption{\textbf{} \\}
    \centering
    \begin{tabular}{cc}
        \toprule
        Parameters & Default Value \\
        \midrule
         Number of agents ($n$) & 1000\\
         Probability of contact ($p$) & 0.003\\
         Rate of Infection ($\beta$) & 0.25\\
         Rate of Recovery ($\gamma$) & 0.2\\
         Gap between testing periods ($T_{G}$) & 1\\
         Number of tests per period ($F$) & 90\\
         Turnaround time for test results ($T_{T}$) & 0\\
         Restriction time for positive results ($T_{R}$)& 6\\
         False Negative rate ($\mu_{-}$)& 0.1\\
         False Positive rate ($\mu_{+}$)& 0.1\\
         \bottomrule
    \end{tabular}
    \label{tab:defparam}
\end{table}

The optimal strategy depends on a wide variety of factors ranging from the cost structure which determines the cost function to the innate environment, disease dynamics and the testing policy itself. 

\begin{figure}[ht]
  \centering
    \resizebox{0.99\textwidth}{!}{\input{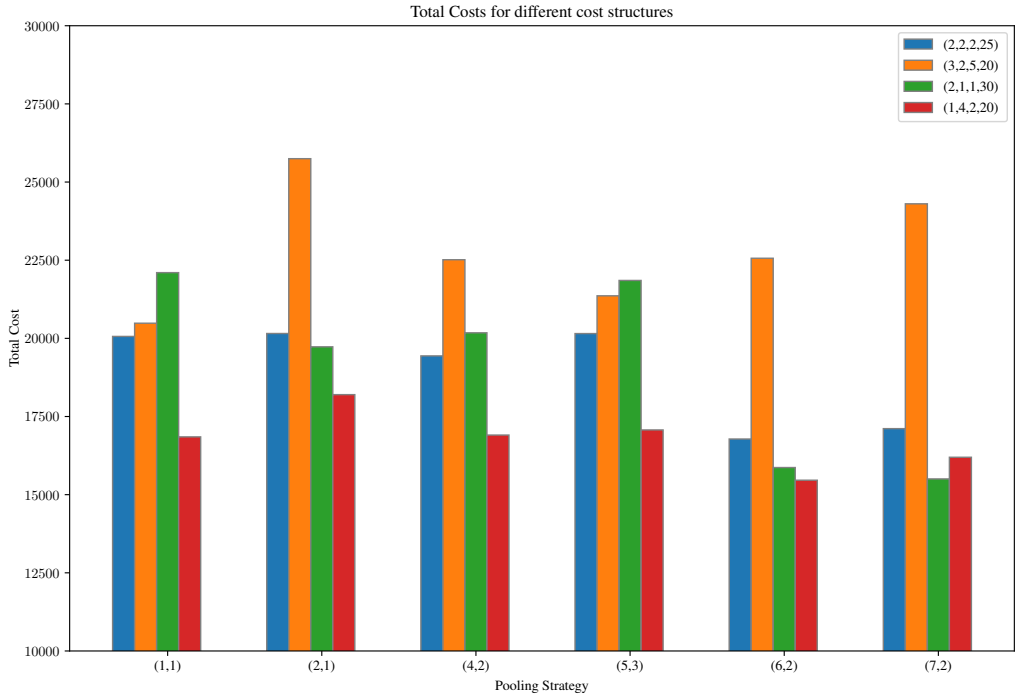}}\\
  \caption{\textbf{Comparison of total cost of different pooling strategies given a cost structure.} This bar-graph shows how the optimal strategy for a community varies with the cost structure. Observe that the pooling strategy (6,2) is optimal for the cost structure (2,2,2,25) and (1,4,2,20) while (1,1) and (7,2) are optimal strategies for (3,2,5,20) and (2,1,1,30) respectively.}
  \label{fig:bargraph}
\end{figure}

\begin{figure}[t!]
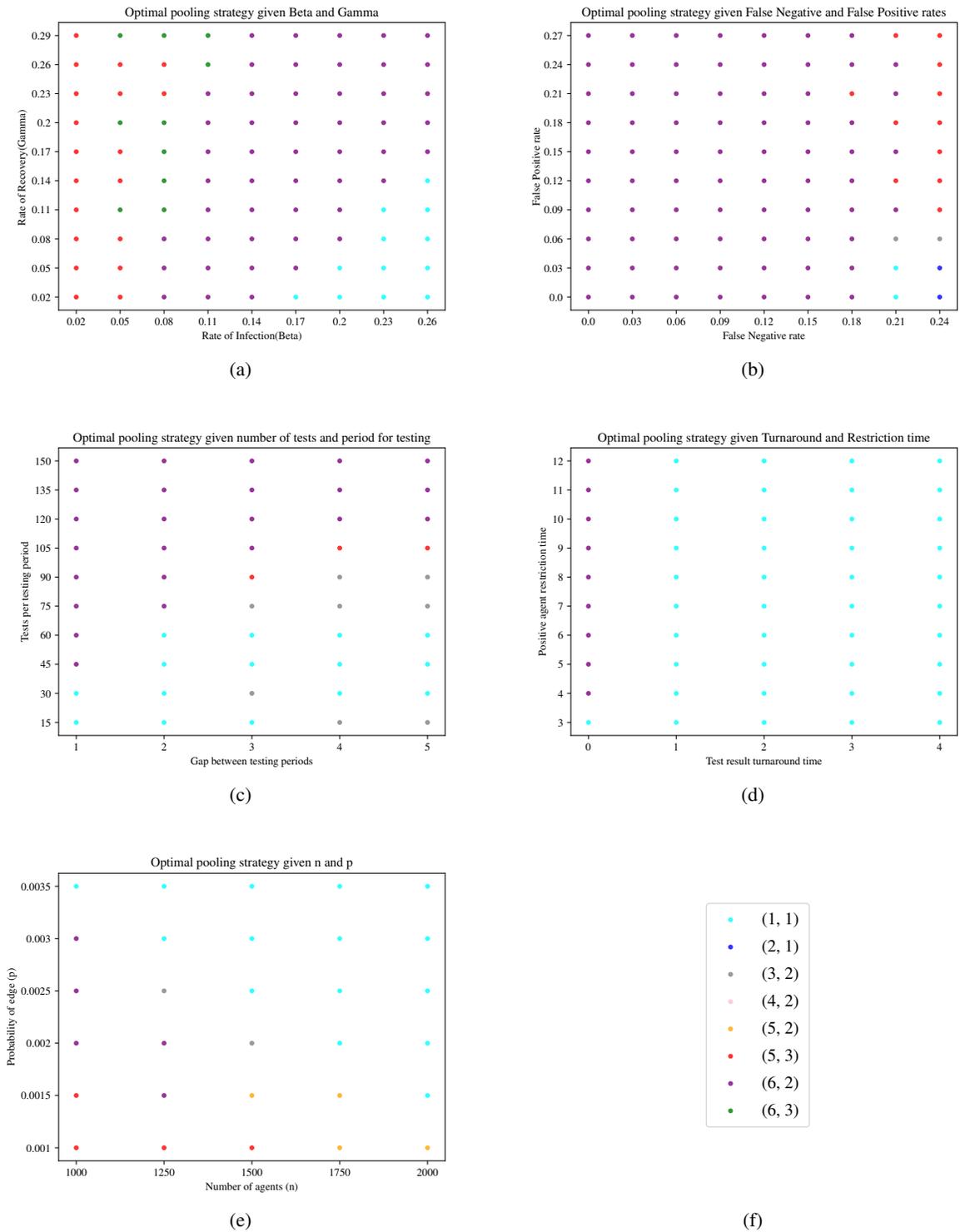

  \centering
    \subfigure[]{\resizebox{0.49\textwidth}{!}{\input{Figures/fig_beta_gamma}}} 
    \subfigure[]{\resizebox{0.49\textwidth}{!}{\input{Figures/fig_fn_fp}}}\\
    \subfigure[]{\resizebox{0.49\textwidth}{!}{\input{Figures/fig_gap_tests}}}
    \subfigure[]{\resizebox{0.49\textwidth}{!}{\input{Figures/fig_turnaround_restriction}}}\\
    \subfigure[]{\resizebox{0.49\textwidth}{!}{\input{Figures/fig_np}}}
    \subfigure[]{\resizebox{0.49\textwidth}{!}{\input{Figures/legend}}}\\
  \caption{\textbf{Optimal pooling strategies varying parameters.} Subfigures (a)-(e) show the optimal strategies varying parameters shown in Table \ref{tab:defparam}. Subfigure (f) is the Legend for plots (a)-(e).}
  \label{fig:scatter}
\end{figure}

We consider a standard set of parameters as shown in Table \ref{tab:defparam}. Figure \ref{fig:bargraph} shows how changes to the cost structure can drastically change the optimal pooling strategy. Furthermore, in Figure \ref{fig:scatter} we vary the other parameters off their default values to see the formations of regions where different strategies are optimal. Each of the subplots varies only two of the ten parameters stated in Table \ref{tab:defparam}. We can see how small changes in some parameters calls for a completely different optimal strategy. \\

When the disease spreads too quickly (large $\beta$), recovers too slowly (small $\gamma$), long time between tests (large gap between testing periods), small number of tests, too many interactions, notice that the pooling strategy (1,1) is most optimal. This is due to quarantining the least individuals in a situation where the disease cannot be controlled. At a more subtle analysis one may find that conducting no tests at all in such situations may be optimal. But conducting no test is highly impractical as it is the only way to derive information about the epidemic all the way from the dynamics of infection and recovery, to the measure its threat level of emerging variants. \\

This analysis brings forth the idea of a personalised strategy for every community. Let alone two similar communities around the world, neighbouring communities may not mandate the same strategy.

\section{Taking a Decision}

We thus come to the critical juncture of actually taking the decision. A community cannot just optimise the cost function as there are real world logistic, supply, infrastructural (Ex : availability of hospital beds), regulatory, medical (Ex : availability of medication) and ethical constraints in place. So the goal is to optimise within the bounds of the constraints. Furthermore there are additional policies and interventions in place which affect the disease spread and also population behaviour, These include vaccination, lockdown, contact tracing etc. - just to name a few. The real world complexity further requires consideration of population compliance and deception. [10, 11] Furthermore all individuals are not homogeneous and their vulnerability to the disease and behaviour vary vastly [12]. \\

\begin{figure}[ht]
  \begin{center}
    \subfigure[]{\includegraphics[scale=0.33]{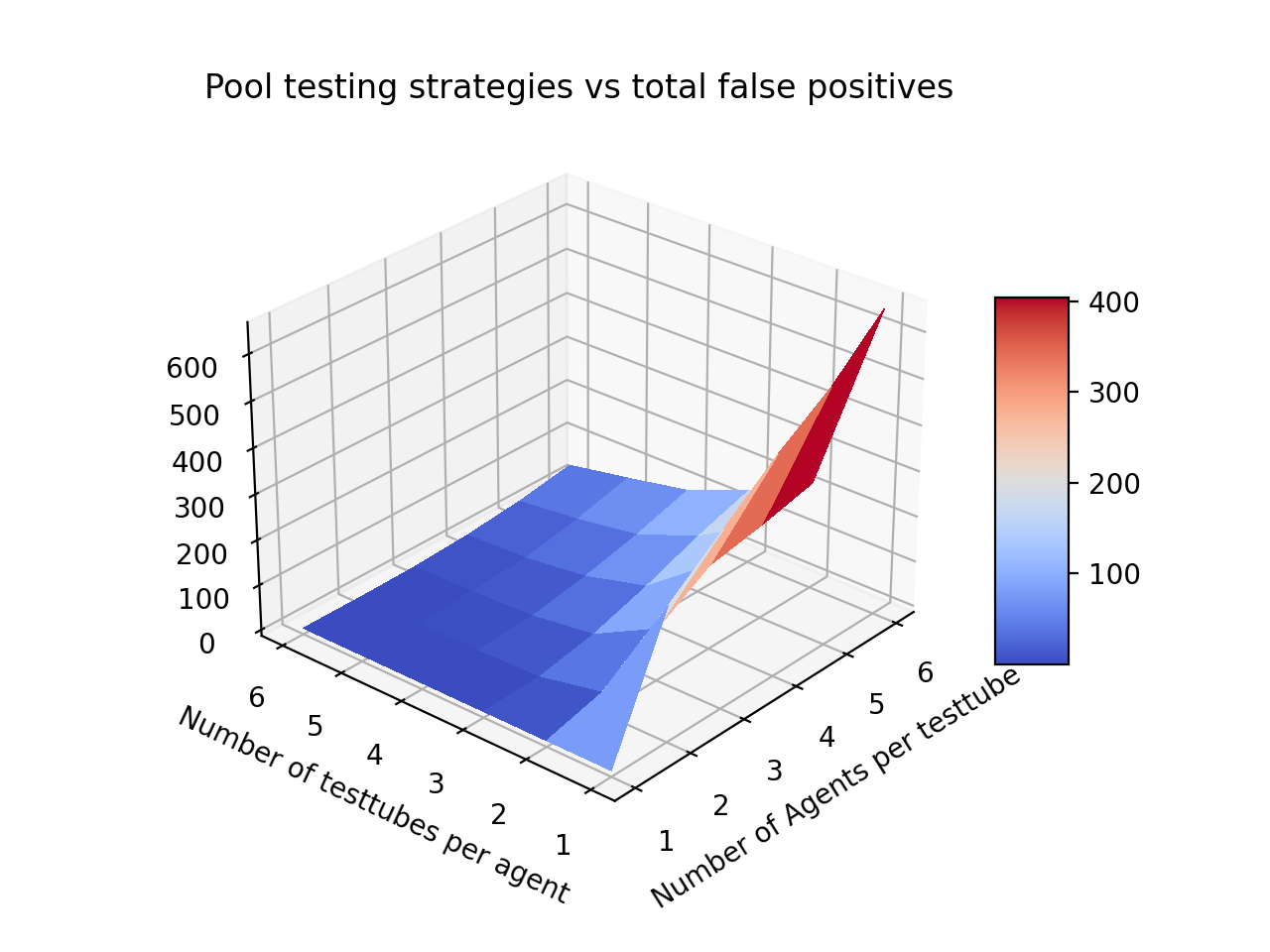}}
    \subfigure[]{\includegraphics[scale=0.33]{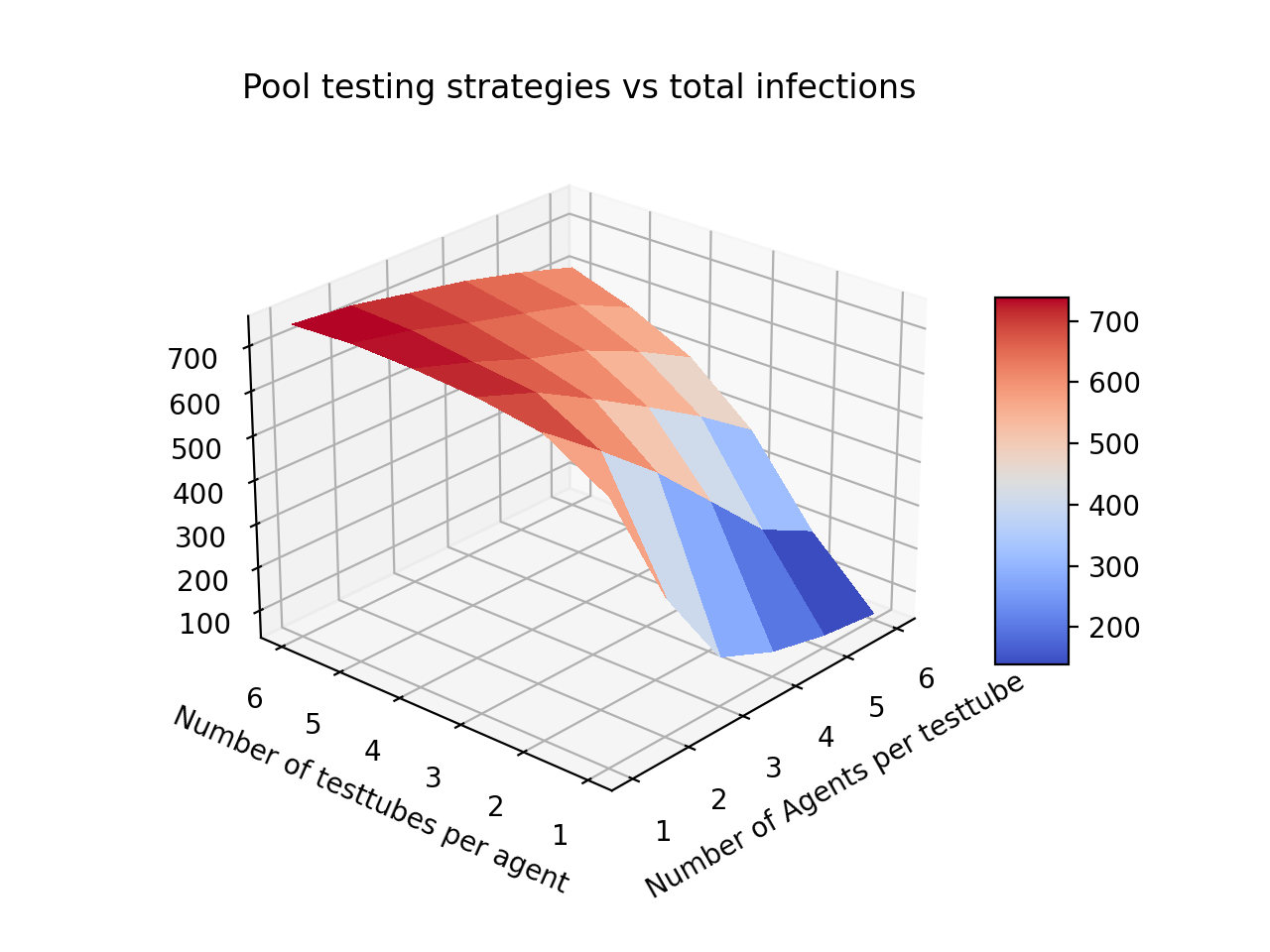}}
    \subfigure[]{\includegraphics[scale=0.33]{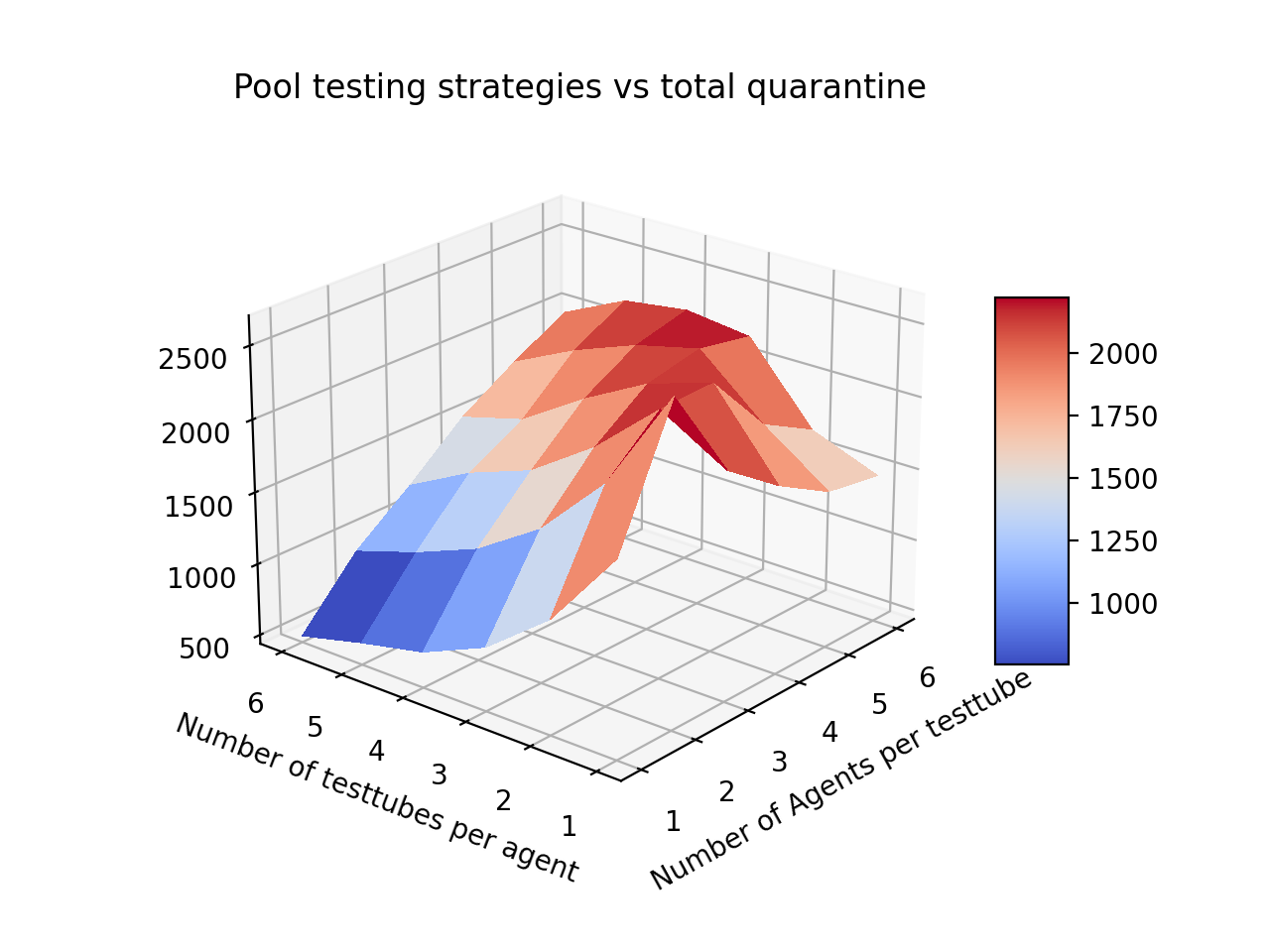}}
  \end{center}
  \caption{\textbf{Visualising tradeoff similar to Figure \ref{fig:3D_plot1}}. The simulations have been run on a G(1000,0.003) ER graph with $\beta=0.25$, $\gamma=0.2$. Subfigures (a),(b) and (c) represent surface plots of total infections, total false positives and cumulative quarantined days across all agents given a 2-dimensional pool testing strategy respectively. The sub figures depict the tradeoff of different quantities for a given strategy.}
  \label{fig:3D_plot2}
\end{figure}

Even after getting past all of this, decision making is not straightforward. It is plagued by a variety of paradoxical and counter-intuitive phenomena. These range from economic and behavioural incongruities like Jevon's Paradox [13], Goodhart's Law [14] to more structural anomalies like Braess Paradox [15] and Parrondo's Paradox [16]. These all have been studied in the epidemic context with very interesting results. In this regard, we shall see an interesting experiment as shown in Figure \ref{fig:3D_plot2}. Observe the non monotonic behaviour in total quarantined days when compared to the pooling strategy. This situation occurs because a pooling strategy with a large $m/k$ ratio essentially quarantines a large number of people early on which borderline converges towards directly locking down the population. This observation raises the dilemma of whether testing is the one which is beneficial or the large restrictions caused by it.\\

Some strategies can be changed real-time according to the situation at hand. But community's have to make an informed decision when it comes to one time decisions like choosing a testing machine or a vaccine for the entire community. Furthermore, these policies interact in complex ways thus making it arduous to choose a policy.\\

Getting all this information and modelling rigorously is a tedious task. In this consideration we have developed an epidemic simulation framework called Episimmer[17] to compare and contrast various strategies. Its high flexibility allows one to experiment various complex strategies with or without complete data. It can help communities like schools and colleges discover and hone the opportunities and optimizations they could make to their epidemic policy. From the most simple decisions(Which days to be online or offline) to more complex strategies(What restrictions should I put on library use?, How many times should I test?, Whom do I test?), Episimmer is a highly promising tool for the job.\\

We have thus shown a versatile random testing strategy and how to optimize a policy at the community level. This calls for a more practical method of moving away from the general approach to policy or testing research towards a more personalised approach. This approach is currently being tried out by HealthBadge with pilot projects in college campuses in India, thus, showing the real world feasibility of such a system.

\section*{Acknowledgments}
We thank RxCovea (\url{https://rxcovea.org/}) a worldwide multidisciplinary research  for useful comments and encouragements. Additionally. we acknowledge HealthBadge, Inc. (\url{https://www.healthbadge.org/}) for its MVPs and pilot projects that demonstrated the real world feasibility and practical application of this idea.

\section*{References}
\begin{enumerate}[label={[\arabic*]}]
    \item Robert Dorfman. The detection of defective members of large populations.The Annals of Mathematical Statistics,14(4):436–440, 1943.
    \item Renato Millioni and Cinzia Mortarino.  Test groups, not individuals:  A review of the pooling approaches forsars-cov-2 diagnosis.Diagnostics, 11(1):68, 2021.
    \item Emilia Vynnycky.An introduction to infectious disease modelling. Oxford University Press, Oxford, 2010.
    \item Jaya Garg, Vikramjeet Singh, Pranshu Pandey, Ashish Verma, Manodeep Sen, Anupam Das, and Jyotsna Agarwal.Evaluation of sample pooling for diagnosis of covid-19 by real time-pcr:  A resource-saving combat strategy.Journal of medical virology, 93(3):1526–1531, 2021.
    \item Khai Lone Lim, Nur Alia Johari, Siew Tung Wong, Loke Tim Khaw, Boon Keat Tan, Kok Keong Chan, Shew FungWong, Wan Ling Elaine Chan, Nurul Hanis Ramzi, Patricia Kim Chooi Lim, et al. A novel strategy for communityscreening of sars-cov-2 (covid-19): Sample pooling method.PloS one, 15(8):e0238417, 2020.
    \item Catherine A Hogan, Malaya K Sahoo, and Benjamin A Pinsky. Sample pooling as a strategy to detect communitytransmission of sars-cov-2.Jama, 323(19):1967–1969, 2020.
    \item Yujia Lu and Kazunori Hayashi. A new pool control method for boolean compressed sensing based adaptive grouptesting.  In2017 Asia-Pacific Signal and Information Processing Association Annual Summit and Conference(APSIPA ASC), pages 994–999, 2017.
    \item Sabyasachi Ghosh, Rishi Agarwal, Mohammad Ali Rehan, Shreya Pathak, Pratyush Agarwal, Yash Gupta, SarthakConsul, Nimay Gupta, Ritika Ritika, Ritesh Goenka, et al. A compressed sensing approach to pooled rt-pcr testingfor covid-19 detection.IEEE Open Journal of Signal Processing, 2021.
    \item Alexander Barg and Arya Mazumdar.  Group testing schemes from codes and designs.IEEE Transactions onInformation Theory, 63(11):7131–7141, 2017.
    \item Alisha Arora, Amrit Kumar Jha, Priya Alat, and Sitanshu Sekhar Das.  Understanding coronaphobia.AsianJournal of Psychiatry, 54:102384, December 2020.
    \item Chinwe U Nnama-Okechukwu, Ngozi E Chukwu, and Chiamaka N Nkechukwu. Covid-19 in nigeria: Knowledgeand compliance with preventive measures.
    \item Eduardo Muñiz-Diaz,  Jaume Llopis,  Rafael Parra,  Imma Roig,  Gonzalo Ferrer,  Joan Grifols,  Anna Millán,Gabriela Ene, Laia Ramiro, Laura Maglio, and et al.  Relationship between the abo blood group and covid-19susceptibility, severity and mortality in two cohorts of patients.Blood Transfusion, January 2021.
    \item Salvador Pueyo. Jevons’ paradox and a tax on aviation to prevent the next pandemic. 2020.
    \item Robert Hancke. Goodhart’s law and the dark side of herd immunity.LSE Covid 19 Blog, 2020.
    \item Hai-Feng Zhang, Zimo Yang, Zhi-Xi Wu, Bing-Hong Wang, and Tao Zhou. Braess’s paradox in epidemic game:better condition results in less payoff.Scientific reports, 3(1):1–8, 2013.
    \item Kang Hao Cheong, Tao Wen, and Joel Weijia Lai. Relieving cost of epidemic by parrondo’s paradox: a covid-19case study.Advanced Science, 7(24):2002324, 2020.
    \item healthbadge/episimmer at deployment.https://github.com/healthbadge/episimmer/tree/deployment.(Accessed on 07/19/2021).

\end{enumerate}

\end{document}